\documentstyle[prl,aps]{revtex}
\begin{document}
\title
{The Mass of the Oppenheimer-Snyder Black Hole}

\author{Abhas Mitra}
\address{Theoretical Physics Division, Bhabha Atomic Research Center,\\
Mumbai-400085, India\\ E-mail: amitra@apsara.barc.ernet.in}


\maketitle

\begin{abstract}
The only instance when the General  Relativistic (GTR) collapse
equations have been solved (almost) exactly to explicitly find the metric
coefficients is the case of a homogeneous spherical dust (
 Oppenheimer and Snyder in 1939 in Phy. Rev.  56,
455).  The equation (37) of their paper showed the formation of an event
horizon for a collapsing homogeneous dust ball of mass $M$ in that the
circumference radius the outermost surface, $r_b\rightarrow r_0= 2GM/
c^2$ in a proper time $\tau_{0} \propto r_{0}^{-1/2}$ in the limit of
large Schrarzschild time $t \rightarrow \infty$. But Eq.(37) was
approximated from the Eq. (36) whose essential character is ($t\sim
r_0 \ln{\sqrt y+1\over \sqrt y-1}$) where, at the boundary of the star $y=r_b/
r_0 = r_b c^2/2G M$. And since the argument of a logarithmic function
can not be negative, one must have $y \ge 1$ or $ 2GM/r_b c^2 \le 1$. This
shows that, atleast, in this case (i) trapped surfaces are not formed,
(ii) if the collapse indeed proceeds upto $r=0$, we must have $M=0$, and
(iii) proper time taken for collapse $\tau \rightarrow \infty$. Thus, the
gravitational mass of OS black holes are unique and equal to zero.

\end{abstract}


One of the oldest and most fundamental problem of physics and astrophysics
is that of gravitational collapse, and, specifically, that of the ultimate
fate of a sufficiently massive collapsing body. Most of the astrophysical
objects that we know of, viz. galaxies, stars, White Dwarfs (WD), Neutron
Stars (NS),  in a broad sense, result from gravitational collapse. 
It is well known that the essential concept of BHs, in a primitive form, was born in
the cradle of Newtonian gravitation\cite{1}. In Newtonian gravity, the mass of a
collapsing gas cloud is constant even if it is radiating, and equal to
its baryonic mass $M_i=M_f=M_0$. Thus as $r$ decreases, natuarally, the value
of $M/r=M_0/r$ steadily increases and at a certain stage one would have
$2GM/r=c^2$ when the ``escape velocity'' from the surface of the fluid
becomes equal to the speed of light $c$ (henceforth $G=c=1$). Soon afterwards, a Newtonian BH
would be born.

And
in the context of classical General Theory of Relativity (GTR), it is believed
that the ultimate fate of sufficiently massive bodies is collapse to a
Black Hole (BH).
In contrast to Newtonian gravity, here, the gravitational mass of a radiating
fluid constantly decrease and therefore unlike the Newtonian case, one can
not predict with real confidence the actual value of $M_f$ when one would
have $2M_f/r =1$. Given an initial gravitational mass $M_i$, the three
quantities, $M_i$, $M_f$ and $M_0$ are not connected amongst themselves by
means of any fundamental constants or by any basic physical principles.
Thus, in a strict sense, one reqires to solve the Einstein equations for
the collapsing fluid for realistic and ever evolving equation of state
(EOS) and radiation transport properties. Unfortunately, the reality is
that even when one does away with the EOS by assuming the fluid to be a
dust whose pressure is zero everywhere including the center, there is no
unique solution to the problem. Depending upon the initial conditions
like density distributions and assumptions, like self-similarity, adopted,
one may find either a BH or a ``naked singularity''\cite{2}.

It is only when the dust is assumed to be homogeneous, a unique solution
can be found\cite{3}. Homogeneous or inhomogeneous, all dust balls have a very
special property: they have no internal energy and they can not radiate.
Consequuenly, the graviational mass of a dust remains constant during the
collapse process, which is, essentially, a Newtonian property. And if the
dust collapse starts from a state of infinite dilution at $r=r_\infty =\infty$,
the graviational mass of a dust must be equal to its baryonic mass
$M=M_0$, another Newtonian property. Having made these remarks, we would
now proceed to self-consistently analyze the pioneering work of
Oppenheimer and Snyder (OS) to determine the mass of the BH whose
formation it suggested.

Since the OS solutions are the only (asymptotic and near) exact solutions for
the GTR collapse, and are believed to explicitly show the formation of
an ``event horizon'' (EH), it is
extremely important to critically reexamine them.
The study of any collapse problem becomes more tractable if one uses the
comoving coordinates which are free from any kind of ``coordinate
singularities''. By definition, for a given fluid element, the comoving
coordinate $R$ is fixed. The most natural choice for $R$ is the number
of baryons within a given mass shell $N(R)$ or any number proportional to it.
The comoving time  $\tau$ is of course the time recordrd by a clock attached to a
fluid element at $R=R$. Since a dust is in perennial free fall, comoving
time is synonymous with ``proper time'' and, therefore, the dust metric is
\begin{equation}
ds^2=d\tau^2 -e^{\bar \omega} dR^2 - e^\omega(d\theta^2 +\sin^2 \theta d\phi^2)
\end{equation}
It turns out that, for any spherically symmetric metric, the angular part
is the same and we have
\begin{equation}
e^\omega= r^2
\end{equation}
where $r$ is the invariant circumference radius, and this was the Eq. (27) in the OS paper.
OS tried to solve the  collapse equation by using the metric (2); and by
skipping several intermediate equations, we shall focus attention on the
solutions they oobtained. The Eq. (23-27) of their paper essentially leads
to a relationship between $\tau$ and $r$:
\begin{equation}
\tau ={2\over 3}{R^{3/2} -r^{3/2}\over (R/R_b)^{3/2} r_0^{1/2}}; \qquad
R\le R_b
\end{equation}
By transposing, the foregoing equation yields
\begin{equation}
{r\over R} =\left(1-{3\over 2} {r_0^{1/2} \tau\over
R_b^{3/2}}\right)^{2/3}; \qquad R\le R_b
\end{equation}
Apart from the the comoving coordinate system, there is another useful
coordinate system, the Schwarzschild system
\begin{equation}
ds^2 = e^\nu dt^2 -e^\lambda dr^2 -r^2 (d\theta^2 +\sin^2 \theta d\phi^2)
\end{equation}
and, in principle, it should be possible to work out the collapse (or
any) problem in this coordinate system. The time $t$ appearing in the
Schwarzschild cooordinate system is the proper time of a distant inertial
observer. Further, by definition, the ``comoving'' coordinates can not be
extended beyond the boundary of the fluid. Thus from either
consideratiion, the external solutions must be
expressed in terms of Schwarzschild system. OS obtained a general form of
the Schwarzschild metric coefficients which involved derivatives, ${\dot
r} = {\partial r/\partial \tau}$ and ${\dot t}={\partial t/\partial \tau}$ :
\begin{equation}
-g_{rr} = e^\lambda =(1-{\dot r}^2)^{-1}
\end{equation}
and
\begin{equation}
g_{tt} =e^\nu =  {\dot t}^{-2} (1- {\dot r})^2
\end{equation}
Again by skipping few intermediate steps, we note that obtained a general
relationship between the coordinate time $t$ and coordinate radius $R$
ifor the region inside the collapsing body:
\begin{equation}
t = {2\over3} r_0 (R_b^{3/2} -r_0^{3/2} y^{3/2}) -2 r_0 y^{1/2} +
r_0 \ln {y^{1/2} +1 \over y^{1/2} -1}
\end{equation}
where
\begin{equation}
y \equiv{1\over 2} \left[ (R/R_b)^2 -1\right] + {R_b r\over r_0  R}
\end{equation}
It is the above Eq. (8) which corresponds to Eq. (36) in the OS paper,
and, for very large $t$, it attains the form
\begin{equation}
t = r_0 \ln {y^{1/2} +1 \over y^{1/2} -1}
\end{equation}
Now we trace some of the intermediate steps used by OS and not contained
in their paper. By using the simple relation $\ln(m/n) = -\ln(n/m)$, one
may rewrite the above equation as
\begin{equation}
t =- r_0 \ln {y^{1/2} -1 \over y^{1/2} +1}
\end{equation}
In the limit of large $t$ and $y\rightarrow 1$, the above equation becomes
\begin{equation}
t = r_0 \ln \left({y-1\over 4}\right)
\end{equation}
Or,
\begin{equation}
y-1 = 4 e^{-t/r_0}
\end{equation}
However, OS  overlooked the numerical factor of ``4'' in
their exercise. From Eqs. (10) and (14), we find that
\begin{equation}
y-1 = {1\over 2} \left[(R/R_b)^2 -3\right] + {R_b r\over r_0 R} = 4e^{-t/r_0}
\end{equation}
The following equation obtained from the foregoing one is used to 
eliminate $r$ from relevant equations
\begin{equation}
{R_b r\over r_0 R} = 4e^{-t/r_0} - {1\over 2} \left[(R/R_b)^2 -3\right]
\end{equation}
And using Eqs. (15) and (16) into (13) we obtain
\begin{equation}
t \sim -r_0 \ln \left\{ {1\over 8}\left[ \left({R \over R_b}\right)^2 -3\right] + {R_b
\over 4 r_0} \left( 1- {3 r_0^{1/2} \tau \over 2 R_b^{3/2}}\right)^{2/3}\right\}
\end{equation}
However, the corresponding equation in the OS paper (Eq.[37]) contained
two small errors:
\begin{equation}
t \sim -r_0 \ln{ \left\{ {1\over 2}\left[ \left({R \over R_b}\right)^2 -
3\right] + {R_b
\over  2r_0} \left( 1- {3 r_0^{1/2} \tau \over 2 R_b^2}\right)^{2/3}\right\}}
\end{equation}
While the first error is a genuine one; which is due to the omission of the
factor 4, the second one is a typographical error : the power of $R_b$ in
the numerator of the last term of this equation should be $3/2$ and
not $2$.
From the foregoing equation, they concluded that, ``for a fixed value of $r$
as $t$ tends toward infinity, $\tau$ tends to a finite limit, which
increases with $r$''. The fact that $\tau$ is finite at $r=r_0$ or $r=0$
is evident from Eq. (4) too (provided $r_b \neq \infty$ and $r_0 >0$). This was
essentially the idea behind the occurrence of an Event Horizon (EH).

By differtiating Eq. (16) with respect to $\tau$, we obtain
\begin{equation}
{\dot t}= {e^{t/r_0}\over 4} \left({r_0 R\over r R_b}\right)^{1/2}
\end{equation}
Similarly differentiating Eq. (15), and using the above result, we obtain
\begin{equation}
{\dot r} = {R^{3/2} r_0^{1/2}\over R_b^{3/2} r^{1/2}}
\end{equation}
Now using Eqs. (15), (19) and (20) in Eqs. (7-8), we obtain
\begin{equation}
e^{-\lambda} = 1 -(R/R_b)^2 \left\{4 e^{-t/r_0} +{1\over 2} \left[3 - (R/R_b)^2\right]\right\}^{-1}
\end{equation}
and
\begin{equation}
e^\nu = e^{\lambda- 2 t/r_0}
\left\{ {e^{- t/r_0}\over 4} + {1\over 8} \left[3-(R/R_b)^2 \right]\right\}
\end{equation}
However, in the paper of OS, these two foregoing equations appear (Eqs.
[38-39]) in a
slightly erroneous form because of the omission of the nemerical factor of
4
in Eq. (13):
\begin{equation}
e^{-\lambda} = 1 -(R/R_b)^2 \left\{ e^{-t/r_0} +{1\over 2} \left[3 - (R/R_b)^2\right]\right\}^{-1}
\end{equation}
and
\begin{equation}
e^\nu = e^{\lambda- 2 t/r_0}
\left\{ e^{- t/r_0} + {1\over 2} \left[3-(R/R_b)^2 \right]\right\}
\end{equation}
Note these equations were obtained by eliminating
$r$ and the
 $t\rightarrow \infty$ limit covers both the
$r\rightarrow r_0$ limit as well as the further $r\rightarrow 0$ limit.
Now, if we recall that the comoving coordinates $R$ and $R_b$ are fixed,
and, when there is a total collapse to a physical point at $r=0$, the
{\em metric coefficients must blow up irrespective of the value of} $R$ (i.e,
the the label of a given mass shell). But this {\em does not happen} for the
solutions obtained by OS! OS correctly 
pointed out that, it is only the outer boundary ($R_b$) for which
the (one of the) metric  coefficients assume the desired form : $e^\lambda
\rightarrow \infty$.  ``For $R$ equal to $R_b$, $e^\lambda$ tends to
infinity like $e^{t/r_0}$ as $t$ approaches infinity.
 But for any interior
point, the limitting values are different, and infact, $e^\lambda =
finite$ even when the collapse is supposed to be complete! OS noted ``For
$R$ less than $R_b$, $e^\lambda$ tends to a finite limit as $t$ tends to
infinity'' {\em without trying to find why it is so}.

And on the other hand, when $e^\lambda =finite$, one sees that $e^\nu
\rightarrow 0$. But, for the external boundary, for which $e^\lambda
\rightarrow \infty$, it can not be predicted whether $e^\nu$ indeed
approaches zero (these limits remain unchanged even when one ignores the numerical factor
of 4).  They admitted this problem while they wrote ``Also for $r \le
r_0$, $\nu$ tends to minus infinity''. This $\nu
\rightarrow -\infty$ limit would correspond to the singularity
$R\rightarrow 0$.  However they
 did not ponder why for $R=R_b$, $e^\nu $ does not behave in the desired manner.

Specifically the fact that for $R <R_b$, $e^\lambda$ {\em fails to blow up at the
singularity definitely 
hints that there is some tacit assumption, made in the OS analysis,  which is not realized in
Nature (GTR) or there is a basic fault in the formulation of the problem}.
In our attempt for a possible resolution of this physical anomaly with
regard to the unphysical aspect of OS solutions, we see
from Eq. (8), that,
$t\rightarrow \infty$ if either or both of the two
following conditions are satisfied:
\begin{equation}
r_0 \rightarrow 0; \qquad t\rightarrow \infty
\end{equation}
and
\begin{equation}
y \rightarrow 1 ; \qquad t\rightarrow \infty
\end{equation}
OS {\em implicitly assumed} that $r_0$ is {\em finite}
 and then $y \rightarrow 1$ and
then $y <1$ as $t\rightarrow \infty$
``For $\lambda$ tends to a finite limit for $r \le r_0$ as $t$
approaches infinity, and for $r_b=r_0$ tends to infinity. Also for $r \le
r_0$, $\nu$ tends to to minus infinity.''
But,
while doing so, {\bf they completely overlooked
 the most important feature} of Eq. (8) (their Eq. 36), and Eq. (10)
that in view of the presence of  the $t \sim \ln {y^{1/2} +1\over y^{1/2} -1}$ term,
 in order that $t$ {\bf is definable at all}, one must have
\begin{equation}
y \ge 1
\end{equation}
For an insight into the problem, we first focus attention on the outermost
layer where $y_b = r_b/r_0$, so that the above condition becomes
\begin{equation}
r_b \ge r_0
\end{equation}
This condition tells that  $r_b$ can {\em never  plunge
below} $r_0$:
Thus a careful analysis of the GTR homogeneous dust problem as enunciated
by OS themselves actually tell that trapped surfaces can not be formed
even though one is free to chase the limit $r\rightarrow r_0$.
We may further rewrite this condition as
\begin{equation}
 {r_b\over r_0} \ge 1; \qquad {2
M\over r_b} \le 1
\end{equation}
This means that, if the collapse indeed proceeds upto $r_b=0$, 
the final gravitational mass of the configuration would be
\begin{equation}
M_f (r=0) = 0
\end{equation}
But then, for a dust or any adiabatically evolving fluid
\begin{equation}
M_i = M_f = constant
\end{equation}
Therefore, we must have $M_i = 0$ too. This means that $r_0=0$ in this
case, and then, it may be promptly verified that, irrespective of the
value of $r$, Eqs. (21) and (22) lead to a
unique limiting value for the metric coefficients :
\begin{equation}
e^{-\lambda} \approx 1 - \left(4 e^{-t/r_0} +1\right) \rightarrow 0
\end{equation}
or,
\begin{equation}
e^\lambda \rightarrow \infty
\end{equation}
and
\begin{equation}
e^\nu \approx e^{\lambda - 2t/r_0}  \rightarrow 0
\end{equation}
because $t/r_0$ approaches $\infty$ much faster than $\lambda$, if $r_0=0$.
Thus, technically, the final solutions of OS are correct, except for the
fact {\bf they did not organically incorporate}  the crucial $y \ge
0$  condition in
the collapse equations (here we ignore the missing numerical factor of 4). And all we have done here is to rectify this
colossal  lacunae to fix the value of $r_0=0$.

This conclusion that the work of OS demands $r_0=0$ could have obtained in
a much more direct fashion simply from the definition of $y$ in Eq. (15).
To this effect, we rewrite this equation as
\begin{equation}
y \equiv {1\over 2} (\alpha^2 -1) + { r\over r_0 \alpha}
\end{equation}
Note that during the collapse process $R/R_b\equiv \alpha$ remains fixed for a given
mass shell. Suppose we are considering the collapse of an interior shell
with $\alpha <1$ and $\alpha^2 -1 <0$. Then {\em if $r_0 >0$, the second term on the left hand
side of the foregoing equation can ne made arbitrarily small} as
$r\rightarrow 0$. Therefore $y$ {\em would become negative} as the collapse
progresses if $r_0 >0$. To avoid this, we must have $r/r_0 \ge 1$.

We may recall here that, long ago, Harrison et al.\cite{4} (pp. 75) mentioned that
 spherical gravitational collapse should
come to a {\em decisive end} with $M_f=M^* =0$. And  they termed this
understanding as a ``Theorem'' (without offering a real proof):

``THEOREM 23: Provided that matter does not undergo collapse at the
microscopic level at any stage of compression, then, regardless of all
features of the equation of state - there exists for each fixed number of
baryons a gravitationally collapsed configuration, in which the
mass-energy $M^*$ as sensed externally is zero.''

In a somewhat more realistic way Zeldovich \& Novikov\cite{5} (see pp.
297) discussed the possibility  of having an ultracompact
configuration of degenerate fermions obeying an equation of state $p=e/3$,
where $e$ is the proper internal energy density, with $M\rightarrow 0$.

And it is also well known that the so-called ``naked singularities'' could
be of zero-gravitational mass\cite{6}.

We have already seen that the coordinate time required for collapse to the
event horizon or beyond is $t=\infty$. And, it may be found that the
proper time for collapse of  the outer boundary of a dust ball to the central singularity is\cite{7}
\begin{equation}
\tau =\pi \left({r_\infty\over 8M}\right)^{1/2}
\end{equation}
where, the dust ball is assumed to be at ``rest'' at $r=r_\infty$ at $\tau
=t=0$.
This equation is also obtainable from Eq. (4) provided one chooses $R_b$ in such
a way thay
\begin{equation}
R_b^{3/2} = {\sqrt 2 \pi\over 3} r_\infty^{3/2}
\end{equation}
Since $M=0$ for the OS problem, {\em the proper time for collapse is
infinite}: $\tau =\infty$, and not finite. Physically this means that, at
any finite proper or coordinate time, there is never any OS-black -hole.
The collapse process goes on and on indefinitely as spacetime becomes
infinitely curved near the would be singularity $r=0$. If this picture is
correct, for self-consistency, Nature should not allow the existence of
finite mass BHs. And at this juncture, some readers may argue that, this
is not so. Irrespective of the results of the OS problem, as highlighted
by us in this paper, it may be argued that GTR allows the existence of
arbitray mass Schwarzschild BHs.

In another related  paper, we assumed the existence of a finite mass
BH which is described by Kruskal- Szekeres\cite{8} metric. If this assumption of the
existence of a finite mass BH is indeed allowed by GTR, one would find the
world line of a material partcle to be {\em timelike}, $ds^2 >0$,  everywhere except
probably at the central singularity $r=0$. However, we have found that,
even in the Kruskal-Szekeres metric the metric {\em appears to be null}\cite{9},
$ds^2 =0$, at $R=2M$. If we describe the region interior to the EH by
Lemaitre metric\cite{10}, we have found that, in this case too, the metric becomes
null at the EH. We have further shown that, in the Kruskal-Szekeres metric,
 the metric continues to
be null for $R <2M$ too\cite{9}. And this can be explained only when we realize
that the Event Horizon, itself is the end of spacetime for the free
falling particle. This implies that mass of the BH is actually $M\equiv
0$, in complete agreement with the present {\em exact} analysis. This is
also in complete agreement with the idea of Einstein that Schwarzschild
singularity can not be realized in practice ($\tau =0$)\cite{11}.

\end{document}